\documentstyle[12pt,aasms4,epsf]{article}

\begin{document}

\title{The Blazhko Effect in RR Lyrae stars: Strong observational 
support for the oblique pulsator model in three stars}

\author{C.~Alcock\altaffilmark{1,2}, 
        D.~Alves\altaffilmark{1,3},
      A.C.~Becker\altaffilmark{5},
      D.P.~Bennett\altaffilmark{6},
      K.H.~Cook\altaffilmark{1,2},
      K.C.~Freeman\altaffilmark{4},
        K.~Griest\altaffilmark{2,7},
      D.W.~Kurtz\altaffilmark{8},
      M.J.~Lehner\altaffilmark{2,9},
      S.L.~Marshall\altaffilmark{1,2},
        D.~Minniti\altaffilmark{1,2,10},
      B.A.~Peterson\altaffilmark{5},
      M.R.~Pratt\altaffilmark{2,6,11},
      P.J.~Quinn\altaffilmark{12},
      A.W.~Rodgers\altaffilmark{5,15},
        W.~Sutherland\altaffilmark{13},
        A.~Tomaney\altaffilmark{6},
        T.~Vandehei\altaffilmark{2,8},
      D.L.~Welch\altaffilmark{14}\\
      {\bf (The MACHO Collaboration\altaffilmark{16}) }
      }

\altaffiltext{1}{Lawrence Livermore National Laboratory, Livermore, CA 94550 \\
        E-mail: {\tt alcock, alves, dminniti, kcook, stuart@igpp.llnl.gov}}
 
\altaffiltext{2}{Center for Particle Astrophysics,
        University of California, Berkeley, CA 94720}
 
\altaffiltext{3}{Department of Physics, University of California,
        Davis, CA 95616 }
 
\altaffiltext{4}{Mt.~Stromlo and Siding Spring Observatories, Australian
        National University, Weston Creek, ACT 2611, Australia \\
        E-mail: {\tt kcf,peterson, alex@mso.anu.edu.au}}
 
\altaffiltext{5}{Departments of Astronomy and Physics,
        University of Washington, Seattle, WA 98195 \\
        E-mail: {\tt becker, austin@astro.washington.edu}}

\altaffiltext{6}{Physics Department, University of Notre Dame, Notre 
        Dame, IN 46556 \\
        E-mail: {\tt bennett.27@nd.edu}}
 
\altaffiltext{7}{Department of Physics, University of California,
        San Diego, La Jolla, CA 92093 \\
        E-mail: {\tt kgriest, tvandehei@ucsd.edu }}
 
\altaffiltext{8}{Department of Astronomy, University of Cape Town, Rondebosch
        7701, South Africa \\
        E-mail: {\tt dkurtz@physci.uct.ac.za}}

\altaffiltext{9}{Department of Physics, University of Sheffield,
        Sheffield, S3 7RH, U.K. \\
        E-mail: {\tt M.Lehner@sheffield.ac.uk}}

\altaffiltext{10}{Departmento de Astronomia, P. Universidad Catolica,
        Casilla 104, Santiago 22, Chile}

\altaffiltext{11}{LIGO Project, MIT, Room 20B-145, Cambridge, MA 02139 \\
        E-mail: {\tt mrp@ligo.mit.edu}}
 
\altaffiltext{12}{European Southern Observatory, Karl-Schwarzchild Str. 2,
        D-85748, Garching, Germany \\
        E-mail: {\tt pjq@eso.org}}

\altaffiltext{13}{Department of Physics, University of Oxford, Oxford OX1
        3RH, U.K. \\
        E-mail: {\tt w.sutherland@physics.ox.ac.uk}}
 
\altaffiltext{14}{Dept. of Physics \& Astronomy, McMaster University,
        Hamilton, Ontario, L8S 4M1 Canada \\
        E-mail: {\tt welch@physics.mcmaster.ca}}

\altaffiltext{15}{Deceased.}

\altaffiltext{16}{The alphabetical list of authors on this paper is a 
subset of the MACHO Project Science Team members and Affiliates}

\begin{abstract}
Using the novel data set of the MACHO project, we show that three ``Blazhko 
Effect'' RR Lyrae stars show nearly-pure amplitude modulation of a single 
pulsation mode. This is strong observational evidence that the 
``Oblique Pulsator Model'' is the correct solution to this 90-year-old problem.
\end{abstract}


\section{Introduction}
RR Lyrae stars are giant A and F stars which pulsate in radial modes with 
periods between 0.2 and 1.2 {\sl d} and with peak-to-peak V amplitudes 
between 0.2 and 2 {\sl mag}. Because of their intrinsic brightness, longevity 
and large amplitudes, they are among the most numerous of all the classes of 
variable stars (in catalogues) and are important to several fields of 
astrophysics: as standard candles at the base of the extragalactic distance 
scale; as tracers of galactic dynamics and chemical evolution among the 
older population of stars; as test objects for stellar evolution theory 
of low-mass stars; and as test objects for stellar pulsation theory. The 
RR Lyrae stars are divided into several subclasses: RRc stars have nearly 
sinusoidal light curves; RRab stars have larger amplitude, sawtooth-shaped 
light curves. For pulsation theory and stellar evolution theory the 
RRd class is of particular interest since these stars pulsate simultaneously 
in the fundamental and first overtone radial modes, the ratio of which places 
a strong constraint on the stellar structure. The most intriguing 
subclass consists of RRab stars which show the ``Blazhko Effect''. 
These stars appear to be fundamental mode pulsators with light curves 
which are modulated on a longer timescale, typically tens to hundreds of 
days. The effect was first noticed by \markcite{Blazhko} Blazhko (1907) for 
the circumpolar RR Lyrae star RW Dra, and it is found in about 30\% of all 
RR Lyrae stars, including the prototype of the class, RR Lyrae itself, which 
was been extensively studied by \markcite{Shapley} Shapley (1916), 
\markcite{Walraven} Walraven (1949), and \markcite{Prestonzz} Preston, 
Smak \& Paczynski (1965). There have been many models put forth for the 
Blazhko Effect. Here we present strong evidence that the oblique pulsator 
model, in which an oblique, global magnetic field modulates the observed 
amplitude with the rotation of the star, is the correct model. Our dataset 
is novel: it has been gathered to study a completely different problem, 
and our data-sampling would previously have incorrectly been considered 
inappropriate to the solution of the Blazhko problem.

The various models that have been proposed to explain the Blazhko Effect are 
listed by \markcite{Stellingwerf} Stellingwerf (1976) and by \markcite{Smith}
Smith (1995) in his monograph (see p. 109) on the RR Lyrae stars. Several of 
the models concentrate on beating of pulsation modes, but they suffer from the 
fact that most, or all, of the Blazhko stars are pulsating in the fundamental 
radial mode. It is necessary to resort to special pleading for non-adiabatic 
effects, non-linear hydrodynamics, tidal resonances, resonances between the 
fundamental radial mode and a higher overtone, or resonances induced by 
(unobserved) non-radial modes for any beating mechanism to be viable. This is 
because the beat periods are long, and there are no pulsation modes with 
frequencies near the fundamental radial mode frequency for adiabatic, 
spherical, non-magnetic single stars. 

Another class of proposed solutions to the problem involves magnetic fields - 
either modulation of the pulsation amplitude caused a magnetic cycle similar 
to that of the sun \markcite{Detre} (Detre \& Szeidl 1973) or oblique pulsation 
similar to that of the rapidly oscillating Ap stars (\markcite{Kurtz-1990} 
Kurtz 1990). This latter model suggests that the Blazhko stars have rigid 
oblique magnetic fields which modulate their pulsation amplitudes with the 
rotation periods of the stars. This idea was vaguely suggested by 
\markcite{Balazs} Bal\'azs (1959), \markcite{Detre} Detre \& Szeidl (1973) and 
\markcite{Preston-1964} Preston (1964). Later it was developed in some detail 
by \markcite{Cousens} Cousens (1983), who called his model the ``Obliquely 
Oscillating Magnetic Rotator''. Independently, \markcite{Kurtz-1982} Kurtz 
(1982) proposed the same model for the rapidly oscillating Ap (roAp) stars and 
called it the ``oblique pulsator model''. That model has been highly successful 
and widely used for the roAp stars, hence we will continue with the name here. 
The oblique pulsator model for the roAp stars has been developed in 
considerable detail by \markcite{ST} Shibahashi \& Takata (1993) and 
\markcite{TS-1994} \markcite{TS-1995} Takata \& Shibahashi (1994, 1995) 
who have extended the idea in an advanced form as the explanation of 
the Blazhko Effect in RR Lyrae stars (\markcite{TS-1998} Takata \& 
Shibahashi 1998).

Until recently it has not been possible to distinguish among this plethora of 
theories. A first requirement to do so is for complete mathematical models 
for many stars of the frequencies describing the light curve modulation. 
Not surprisingly, traditional observing methods preclude this. Even 
though 90 years has passed and the problem is important, the observational 
demands are extreme for a single observer, or a small group of observers. 
The problem is well-illustrated by the work of \markcite{Preston-1965} 
Preston, Smak \& Paczynski (1965) who observed RR Lyrae itself on 35 nights. 
This star has a Blazhko period of 41 {\sl d}. Their data mostly cover 
the sharp rise of the lightcurve to its brightness maximum; very little 
data were obtained during the slower descending portions of the lightcurves. 
This was a typical observing technique at the time when frequency analysis 
was done by studying the timing of maxima, the so-called ``O-C method'', and 
amplitude variations were determined from the range of minimum to maximum - 
observationally most easily found in the shorter rise-time portion of the 
non-sinusoidal lightcurve. Observations such as these send modern, 
powerful frequency analysis techniques, such as Fourier analysis, into a 
frenzy! 

What is needed are complete light curves over the pulsation period (about 
12 hours, typically) and the Blazhko period (months, typically). To do this 
means either: 1) a multisite campaign with observers stationed at sites 
around the globe continuously for more than one Blazhko cycle - months of 
work for probably half a dozen observers, all to get data for only a few stars 
at best; or 2) a very long-term observing project from one site to build up 
a sufficient data set, or 3) a revolutionary new way to tackle the problem. 

Remarkably, the second method has shown some success. \markcite{Kovacs} 
Kov\'acs (1995), working with a heroic data set mostly obtained by 
\markcite{Kanyo} Kany\'o(1976) over a period of 6 years, has shown that 
the frequencies of the circumpolar RR Lyrae star RV UMa comprise an 
equally-spaced triplet plus its harmonics. This is close to the 
predicted frequency behavior for the oblique pulsator model. Pure 
amplitude modulation of a magnetically-distorted, fundamental radial 
mode in the oblique pulsator model generates a frequency multiplet 
with spacings exactly equal to the rotation frequency of the star 
(\markcite{TS-1998} Takata \& Shibahashi 1998). 
Pure amplitude modulation also requires equality of the phases of the 
frequency multiplet in the oblique pulsator model, a condition which was 
not examined by Kov\'acs, but which does not precisely obtain for RV UMa. 
Because of the lack of any believable magnetic measurements for the Blazhko 
stars, and because of a perceived need for non-radial pulsation modes, as 
observed in the roAp stars, Kov\'acs was very reserved about the viability 
of the oblique pulsator model explanation of the Blazhko Effect, and, 
for that matter, all the other models, too. To make progress on this 
nearly century-old problem we have utilized the extensive photometric
databases of the microlensing survey known as the MACHO Project.

\section{MACHO Project Observations}

The MACHO (MAssive Compact Halo Object) Project (\markcite{Alcock-1992} 
Alcock {\it et al.} 1992) is an astronomical survey experiment designed 
to obtain multi-epoch, two-color CCD photometry of millions of stars in 
the Large Magellanic Cloud (LMC) (also, in the galactic bulge and Small 
Magellanic Cloud = SMC). The principal goal of the project is to search for 
massive compact objects whose presence between the observer and a background 
source will result in an amplification of received flux caused by gravitational 
lensing. The expected rate of detectable events is very low, requiring large 
numbers of background sources - in this case, LMC stars - to be measured over 
many years. The survey makes use of a dedicated 1.27-m telescope at Mount 
Stromlo, Australia and because of its southerly latitude is able to obtain 
observations of the LMC year-round. This lack of seasonal aliasing is 
one factor which makes these observations so suitable for studying the 
Blazhko effect in RR Lyrae stars. 

The camera built specifically for this project (\markcite{Stubbs} Stubbs 
{\it et al.} 1993) achieves a field-of-view of 0.5 square degrees by imaging 
at prime focus. Observations are obtained in two bandpasses simultaneously, 
using a dichroic beamsplitter to direct the `blue' (approximately 
450-630 {\sl nm}) and `red' (630-760 {\sl nm}) light onto 2$\times$2 
mosaics of 2048$\times$2048 pixel Loral CCD chips. Images are obtained 
and read out simultaneously. The 15 {\sl $\mu$m} pixel size maps to 0.63 
{\sl arcsec} on the sky. The data were reduced using a profile-fitting 
photometry routine known as Sodophot, derived from DoPHOT (\markcite{Schechter}
Schechter, Mateo \& Saha 1993). This implementation employs a single starlist 
generated from frames obtained in good seeing. 

The results reported in this paper comprise only a small fraction of all the 
RR Lyrae stars (Blazhko and non-Blazhko) found by the MACHO Project. In the 
first instance, we sought stars which were likely foreground RR Lyrae stars 
according to \markcite{Alcock-1997} Alcock {\it et al.} (1997). Before 
analysis, lightcurves were brought up to date to maximise the number of 
Blazhko cycles observed. Many more Blazhko RR Lyrae stars have been identified 
(and many remain to be confirmed) based on goodness-of-fit statistics for 
single-cycle periods for variable stars with appropriate magnitudes and 
colors toward the LMC and SMC. 

Typically, the dataset for a given star covers a time-span of about 2000 {\sl d}
and contains 750 photometric measurements (multiple observations of some
fields are obtained on a given night whenever conditions allow). Fainter, 
LMC member stars may have fewer useful epochs of photometry. The output 
photometry contains flags indicating suspicion of errors due to crowding, 
seeing, array defects, and radiation events. Only data free from 
suspected errors were employed for this work. Typical photometric 
uncertainties are in the range 15 to 20 {\sl mmag} (millimagnitudes) or 
1.5 to 2\% in intensity. 

With only one or two points per night, it would not seem at first thought 
to be possible to do a definitive frequency analysis for stars with pulsation 
periods on the order of 12 hours. For equally-spaced data it is well known 
that the frequency solutions from Fourier analysis are ambiguous with equally 
probable solutions occurring above and below the Nyquist frequency, twice 
the Nyquist frequency, etc. This is simply because of beating between the real 
frequencies and the sampling frequency. However, with unequally-spaced data, 
such as the MACHO data, the ambiguities disappear for large enough data sets. 
``Large enough'' is easy to define: it means that the data gathered are 
well-spread over all phases of all frequencies present in the data. These 
criteria are met for many of the Blazhko stars in the MACHO Project dataset.

Some of these stars show pure amplitude modulation of their light curves 
over the Blazhko period. This is as predicted for the oblique pulsator model 
for these stars, hence it gives very strong support for this model. Other 
stars, which will be discussed in future publications, show pure 
frequency modulation - they are probably binary stars where the 
pulsation frequency is Doppler-shifted by the orbital motion. Most stars 
show both amplitude and frequency modulation. Whether this is simply 
a combination of oblique pulsation and orbital motion, or whether there 
are further complexities to the physics of some stars classified 
as Blazhko stars is still to be determined. 

\section{Analysis, Results and Discussion}

We have performed frequency analyses on over 20 RR Lyrae stars from 
the MACHO Project data using standard Discrete Fourier Transforms 
(\markcite{Deeming} Deeming 1975; \markcite{Kurtz-1985} Kurtz 1985). 
Three of the stars show nearly the pure amplitude modulation expected 
from the oblique pulsator model. It can be seen in Figure 1
from the tightness of the rising branch 
that there is little, or no phase modulation in star 82.8410.55 
(our star naming convention is field.tile.sequence - see 
\markcite{Alcock-1992} Alcock {\it et al.} 1992). It can also be seen 
here and in Figure 2 that the amplitude 
modulation mostly affects the maximum of the lightcurve, and has a 
much smaller effect on the minimum. 

Table \ref{freq-anal-82.8410.55} shows the results of our frequency analysis 
for star 82.8410.55. Following the notation of \markcite{Kovacs} Kov\'acs 
(1995): We find that a stationary Fourier-sum with frequencies of the 
form $kf_0$, $kf_0 \pm f_m$, where $k$ = 1, 2, 3, 4, 5, 6, 7; $f_0$ = 
1.932166 $\pm$ 0.000045 {\sl d}$^{-1}$ and $f_m$ = 0.011642 $\pm$ 0.000029 
{\sl d}$^{-1}$ gives a complete fit to the data to a standard deviation 
per observation of 0.028 mag, which is close to our estimated error. We 
have chosen a zeropoint to the timescale for which the phases 
$\varphi(f_0 + f_m)$ = $\varphi(f_0 - f_m)$. For pure amplitude modulation 
the $\varphi(f_0)$ would also be equal to these other phases. It can be 
seen in Table \ref{freq-anal-82.8410.55} that this is nearly the case, 
although the deviation is significant by 10$\sigma$.

All of the above is consistent with the oblique pulsator model as formulated 
by \markcite{TS-1998} Takata \& Shibahashi (1998). In that model an 
oblique, dipole magnetic field modulates the amplitude of the 
fundamental radial mode as a function of rotation. Mathematically, the 
result can be described as a mode with radial and quadrupole 
contributions. However, this does not imply that the mode is 
non-radial. The mode excited is still the radial fundamental mode, 
but it no longer can be described by a single spherical harmonic, 
because of the distorting effect of the magnetic field which suppresses 
the pulsation surface-brightness amplitude at the magnetic equator where 
the photosphere has to cross field lines, while doing little to modify 
the amplitude at the magnetic poles where the motion is along 
the field lines. 

Takata \& Shibahashi predict that a frequency quintuplet will be 
generated by the amplitude modulation, but it is clear from their 
equations and diagrams that for many orientations the outer two 
frequencies of the quintuplet have much lower amplitudes, so it is 
not surprising that we have yet to detect them. That will 
take better signal-to-noise ratio data than we have, or the discovery 
of a star with an appropriate rotational inclination and magnetic 
obliquity. A similar result is seen for the roAp stars where a 
frequency triplet can be seen in several stars, and where very 
high signal-to-noise ratio data show a frequency septuplet in 
HR 3831 (\markcite{Kurtz-1997} Kurtz {\it et al.} 1997).

Similarly, the small, but significant difference in the phase $\varphi(f_0)$ and 
$\varphi(f_0 \pm f_m)$ is not surprising. This same effect is seen in the 
roAp stars, particularly HR 3831. This is explained by a deviation 
from pure dipolar symmetry for the magnetic field - a deviation 
which is known to exist in all Ap stars for which sufficiently high 
accuracy observations are available.  

Our results for star 82.8410.55 show almost pure amplitude modulation with a 
period of 85.9 $\pm$ 0.2 {\sl d}. We conclude that this is strong 
evidence in favor of the oblique pulsator model for this star, and by 
extension for the Blazhko Effect. There is no need to invoke non-radial 
modes as many other Blazhko explanations do. Only the fundamental radial 
mode is excited. Models which suggest more than one excited mode are 
ruled out by the frequencies in Table \ref{freq-anal-82.8410.55}. If 
$f_0 - f_m$, $f_0$ and $f_0 + f_m$ were independent modes, then we 
would expect to see their harmonics, since the pulsation is non-linear. 
We do not see this. Instead, we see the multiples $kf_0$ split by $\pm f_m$, 
as expected in the oblique pulsator model. Notice also that the phases 
of the frequency triplets for the harmonics also show nearly equal 
phases - indicating that the harmonics are amplitude-modulated in phase 
with the fundamental. Independently excited modes would not necessarily 
behave that way. 

\markcite{Kovacs} Kov\'acs (1995) suggested several potential weaknesses 
in the oblique pulsator model: 1) There is no verification of magnetic 
fields in the Blazhko stars; 2) all known Blazhko stars have modulation 
amplitudes in the range 0.3 -- 0.7 {\sl mag}, whereas there should be 
no lower limit, since the amplitude range is a function of aspect angle, 
and all angles are permissible; and 3) the Blazhko stars fit the 
standard period-amplitude relation in their high-amplitude state. We 
note that 1) Takata \& Shibahashi's formulation of the oblique pulsator model 
requires field strengths of about 1 kG. At present there are no observations 
which confirm or refute the presence of magnetic fields of this strength. 
Such observations are strongly desirable. 2) The lower limit to the 
modulation amplitude may be a discovery selection bias, so that lower 
amplitude Blazhko stars are more difficult to detect, hence are not yet 
noted. We agree that conclusive proof that no Blazhko stars exist with 
amplitudes lower than 0.3 {\sl mag} would constitute a strong challenge 
to the oblique pulsator model. Our large dataset should allow us to 
examine this. 3) In Takata \& Shibahashi's formulation of the oblique 
pulsator model the energy of the mode is not affected by the presence of the 
magnetic field, so the amplitude at maximum is equal to that which would obtain 
in the absence of any magnetic field, hence the standard period-amplitude 
relation is expected to fit. 

The last three columns of Table \ref{freq-anal-82.8410.55} give quantitative 
measures which can be tested against more sophisticated predictions of the 
oblique pulsator model when the theory is more advanced, and against 
magnetic observations when they become available. The parameter 
${{A_{+1} + A_{-1}} \over {A_0}}$ depends on the inclination of the 
rotation axis of the star to the line-of-sight and the obliquity of the 
magnetic axis to the rotation axis. If the Blazhko stars have 
approximately dipolar magnetic fields, then these geometrical angles can be 
derived independently from magnetic observations. That 
${{A_{+1} + A_{-1}} \over {A_0}}$ grows for the higher harmonics is a 
consequence of the amplitude modulation shown in Figure 2.
There is no theoretical explanation of 
this yet - this is the first observational description of the effect. 
The same holds true for the phases in the penultimate column of 
Table \ref{freq-anal-82.8410.55}. We predict from the oblique pulsator 
Model that the time of magnetic maximum for star 82.8410.55 will coincide 
with the time of pulsation maximum seen in Figure 2 and given in the 
caption to Table \ref{freq-anal-82.8410.55}.

We have two more stars, AC Men (6.5722.3, also known as HV 924) and 
5.5376.3686, which show pure amplitude modulation. Tables 
\ref{freq-anal-6.5722.3} and \ref{freq-anal-5.5376.3686} show the 
frequency structure for these stars. Figures 3--6
show the phased lightcurves. AC Men even has exact equality of the phases 
$\varphi(f_0)$ = $\varphi(f_0+f_m)$ = $\varphi(f_0-f_m)$. We consider the pure 
amplitude modulation of this, and the nearly pure amplitude modulation of 
stars 82.8410.55 and 5.5376.3686 very strong support for the oblique 
Pulsator Model for the Blazhko effect as formulated by Takata \& 
Shibahashi. There is a great wealth of astrophysical data yet to be 
extracted from our observations and frequency analyses. We have stars 
which show pure frequency modulation, and stars which show a combination 
of frequency and amplitude modulation over their Blazhko cycles. The 
amplitudes and phases of the determined frequencies give detailed 
information about the behavior of the pulsation harmonics. Understanding 
these astrophysically is a great challenge for non-linear pulsation theory. 
A taste of this is shown in Figure 2 where it 
can be seen that the Blazhko amplitude modulation for star 82.8410.55 
occurs mostly at the maximum of the light curve. The modulation at 
minimum is much less pronounced. Within the oblique pulsator model this 
means that the minimum state is much less affected by the orientation of 
the field lines than the maximum state - an interesting observation, and 
an interesting problem. 

To assist other observers in studying these stars, we provide finder
charts from `red' MACHO Project images for 82.8410.55, AC Men (6.5722.3),
and 5.5376.3686 in Figures 7, 8, and 9, respectively.

We can look forward to a flood of more information and interpretation of the 
Blazhko stars as we mine the vast data resources of the MACHO Project. 
Finally, after 90 years, we have found the practical observing technique 
for these stars.

\acknowledgements

Acknowledgements We are very grateful for the skilled support given our 
project by the technical staff at the Mt. Stromlo Observatory. Work 
performed at LLNL is supported by the DOE under contract W7405-ENG-48. 
Work performed by the Centre for Particle Astrophysics on the UC campuses 
is supported in part by the Office of Science and Technology Centres of 
NSF under co-operative agreement AST-8809616. Work performed at MSSSO 
is supported by the Bilateral Science and Technology Program of the 
Australian Department of Industry, Technology and Regional Development. 
KG acknowledges a DOE OJI grant, and CWS and KG thank the Sloan 
Foundation for their support. DLW was a Natural Sciences and Engineering 
Research Council of Canada (NSERC) University Research Fellow during this 
work.


\newpage
\begin{deluxetable}{lrrrrrc}
\footnotesize
\tablecaption{Frequency Solution for 82.8410.55
\label{freq-anal-82.8410.55}} 
\tablewidth{0pt}
\tablehead{
&
\colhead{Frequency} &  
\colhead{A} &
\colhead{$\varphi$} &
\colhead{${{A_{+1} + A_{-1}} \over {A_0}}$} &
\colhead{$\varphi_{+} - \varphi_{-}$} &
\colhead{ ${{\varphi_{+} - \varphi_{-}} \over {\sigma}}$}\\
&
\colhead{{\sl (d$^{-1}$)}}&
\colhead{{\sl (mmag)}}&
\colhead{{\sl (rad)}}&
&
\colhead{ $\varphi - { {\varphi_+ + \varphi_-}\over{2} } $}&
\colhead{ ${\varphi - {{\varphi_+ + \varphi_-} \over {2}}} \over {\sigma}$ }
}
\startdata
$f_0$ - $f_m$& 1.932194& 54.0 $\pm$ 1.5& 2.216 $\pm$ 0.029& & 0.000 $\pm$ 0.042 &0.0\nl
$f_0$& 1.943836& 369.8 $\pm$ 1.5& 1.987 $\pm$ 0.004& 0.28 $\pm$ 0.01& -0.229 $\pm$ 0.021& -10.8\nl
$f_0$ + $f_m$& 1.955478& 50.3 $\pm$ 1.5& 2.216 $\pm$ 0.031& & & \nl
&&&&&&\nl
$2f_0$ - $f_m$& 3.876030& 31.9 $\pm$ 1.5& 1.741 $\pm$ 0.048& & -0.016 $\pm$ 0.071& -0.2\nl
$2f_0$& 3.887672& 149.2 $\pm$ 1.5& 1.545 $\pm$ 0.011& 0.42 $\pm$ 0.02& -0.188 $\pm$ 0.036& -5.3\nl
$2f_0$ + $f_m$& 3.899314& 30.6 $\pm$ 1.5& 1.725 $\pm$ 0.052& & & \nl
&&&&&&\nl
$3f_0$ - $f_m$& 5.819866& 30.8 $\pm$ 1.5& 1.339 $\pm$ 0.051& & 0.105 $\pm$ 0.077& 1.4\nl
$3f_0$& 5.831508& 102.4 $\pm$ 1.5& 1.365 $\pm$ 0.016& 0.56 $\pm$ 0.02& -0.027 $\pm$ 0.039& -0.7\nl
$3f_0$ + $f_m$& 5.843150& 26.7 $\pm$ 1.5& 1.444 $\pm$ 0.058& & & \nl
&&&&&&\nl
$4f_0$ - $f_m$& 7.763702& 24.2 $\pm$ 1.5& 1.067 $\pm$ 0.064& & 0.416 $\pm$ 0.103& 4.0\nl
$4f_0$& 7.775344& 62.6 $\pm$ 1.5& 1.293 $\pm$ 0.025& 0.69 $\pm$ 0.04& 0.018 $\pm$ 0.052& 0.3\nl
$4f_0$ + $f_m$& 7.786986& 19.2 $\pm$ 1.5& 1.483 $\pm$ 0.081& & & \nl
&&&&&&\nl
$5f_0$ - $f_m$& 9.707538& 13.2 $\pm$ 1.5& 0.818 $\pm$ 0.116& & 0.519 $\pm$ 0.159& 3.3\nl
$5f_0$& 9.719180& 37.2 $\pm$ 1.5& 1.191 $\pm$ 0.041& 0.74 $\pm$  0.07& 0.114 $\pm$ 0.081& 1.4\nl
$5f_0$ + $f_m$& 9.730822& 14.2 $\pm$ 1.5& 1.337 $\pm$ 0.109& & & \nl
&&&&&&\nl
$6f_0$ - $f_m$& 11.651374& 11.1 $\pm$ 1.5& 0.481 $\pm$ 0.139& & 0.614 $\pm$ 0.250& 2.5\nl
$6f_0$& 11.663016& 22.4 $\pm$ 1.5& 1.125 $\pm$ 0.069& 0.83 $\pm$  0.11& 0.337 $\pm$ 0.130& 2.6\nl
$6f_0$ + $f_m$& 11.674658& 7.4 $\pm$ 1.5& 1.095 $\pm$ 0.208& & & \nl
&&&&&&\nl
\tablebreak
$7f_0$ - $f_m$& 13.595210& 8.2 $\pm$ 1.5& 0.424 $\pm$ 0.188& & 0.622 $\pm$ 0.293& 2.1\nl
$7f_0$& 13.606852& 9.4 $\pm$ 1.5& 0.858 $\pm$ 0.162& 1.60 $\pm$ 0.35& 0.123 $\pm$ 0.173& 0.7\nl
$7f_0$ + $f_m$& 13.618494& 6.9 $\pm$ 1.5& 1.046 $\pm$ 0.225& & & \nl
\enddata

\tablecomments{The star pulsates with only one frequency, $f_0$ =
1.943836 $\pm$ 0.000006 {\sl d}$^{-1}$ ({\sl P} = 12.346721 $\pm$
0.00038 {\sl hours}). The amplitude of this frequency is modulated
with the Blazhko frequency of $f_m$ = 0.01164 $\pm$ 0.00003 {\sl d}$^{-1}$  
({\sl P}$_{Blazhko}$ = 85.9 $\pm$ 0.2 {\sl d}), which, within the
oblique pulsator model, we equate to the rotation frequency of the
star (or twice the rotation frequency if both pulsation/magnetic
poles are seen from equal aspect). The parameters in this table fit
the relation:
$$mag=A_0+\sum\limits_{k=1}^n {\left\{ \matrix{A_-(k)\cos \left[ {2\pi \left( {kf_0-f_m} 
\right)\left( {t-t_0} \right)+\varphi _-(k)} \right]\hfill\cr
  A_0(k)\cos \left[ {2\pi kf_0\left( {t-t_0} \right)+\varphi (k)} \right]\hfill\cr
  A_+(k)\cos \left[ {2\pi \left( {kf_0+f_m} \right)\left( {t-t_0} \right)+\varphi _+(k)} 
\right]\hfill\cr} \right\}}$$
where $n$ = 7 and $t_0$ = HJD 2449854.8322 for our instrumental
magnitudes. Predicting the quantities in the last three columns,
particularly for the harmonic frequencies, presents a strong challenge
to non-linear pulsation theory.}

\end{deluxetable}

\newpage
\begin{deluxetable}{lrrrrrc}
\footnotesize
\tablecaption{Frequency Solution for AC Men (6.5722.3)
\label{freq-anal-6.5722.3}}
\tablewidth{0pt}
\tablehead{
&
\colhead{Frequency} &
\colhead{A} &
\colhead{$\varphi$} &
\colhead{${{A_{+1} + A_{-1}} \over {A_0}}$} &
\colhead{$\varphi_{+} - \varphi_{-}$} &
\colhead{ ${{\varphi_{+} - \varphi_{-}} \over {\sigma}}$}\\
&
\colhead{{\sl (d$^{-1}$)}}&
\colhead{{\sl (mmag)}}&
\colhead{{\sl (rad)}}&
&
\colhead{ $\varphi - { {\varphi_+ + \varphi_-}\over{2} } $}&
\colhead{ ${\varphi - {{\varphi_+ + \varphi_-} \over {2}}} \over {\sigma}$ }
}
\startdata
$f_0$ - $f_m$&1.798715&39.7 $\pm$ 3.5&2.234 $\pm$ 0.088& &0.00 $\pm$ 0.12& \nl
$f_0$&1.806942&414.8 $\pm$ 3.5&2.195 $\pm$ 0.008&0.19 $\pm$ 0.01&-0.04 $\pm$ 0.06&-0.6\nl
$f_0$ + $f_m$&1.815169&39.5 $\pm$ 3.5&2.234 $\pm$ 0.088& & & \nl
&&&&&&\nl
$2f_0$ - $f_m$&3.605657&7.9 $\pm$ 3.5&1.271 $\pm$ 0.450& &0.92 $\pm$ 0.47&1.9\nl
$2f_0$&3.613884&175.5 $\pm$ 3.5&1.954 $\pm$ 0.020&0.18 $\pm$ 0.03&0.23 $\pm$ 0.24&0.9\nl
$2f_0$ + $f_m$&3.622111&23.7 $\pm$ 3.5&2.187 $\pm$ 0.148& & & \nl
&&&&&&\nl
$3f_0$ - $f_m$&5.412599&19.2 $\pm$ 3.5&1.361 $\pm$ 0.182& &0.78 $\pm$ 0.25&3.0\nl
$3f_0$&5.420826&116.1 $\pm$ 3.5&2.055 $\pm$ 0.030&0.33 $\pm$ 0.04&0.31 $\pm$ 0.13&2.4\nl
$3f_0$ + $f_m$&5.429053&19.5 $\pm$ 3.5&2.137 $\pm$ 0.178& & & \nl
&&&&&&\nl
$4f_0$ - $f_m$&7.219541&15.8 $\pm$ 3.5&1.159 $\pm$ 0.224& &1.30 $\pm$ 0.32&4.0\nl
$4f_0$&7.227768&69.8 $\pm$ 3.5&2.220 $\pm$ 0.050&0.44 $\pm$ 0.07&0.41 $\pm$ 0.16&2.5\nl
$4f_0$ + $f_m$&7.235995&14.8 $\pm$ 3.5&2.463 $\pm$ 0.235& & & \nl
&&&&&&\nl
$5f_0$ - $f_m$&9.026483&10.1 $\pm$ 3.5&2.026 $\pm$ 0.346& &0.40 $\pm$ 0.40&1.0\nl
$5f_0$&9.034710&44.8 $\pm$ 3.5&2.221 $\pm$ 0.078&0.60 $\pm$ 0.12&0.00 $\pm$0.21&0.0\nl
$5f_0$ + $f_m$&9.042937&16.8 $\pm$ 3.5&2.422 $\pm$ 0.206& & & \nl
&&&&&&\nl
$6f_0$ - $f_m$&10.833425&9.7 $\pm$ 3.5&2.422 $\pm$ 0.360& &0.06 $\pm$ 0.47&0.1\nl
$6f_0$&10.841652&31.3 $\pm$ 3.5&2.404 $\pm$ 0.112&0.69 $\pm$ 0.18&-0.05 $\pm$ 0.25&-0.2\nl
$6f_0$ + $f_m$&10.849879&11.8 $\pm$ 3.5&2.487 $\pm$ 0.296& & & \nl
&&&&&&\nl
$7f_0$ - $f_m$&12.640367&8.3 $\pm$ 3.5&2.288 $\pm$ 0.418& &0.29 $\pm$ 0.55&0.5\nl
$7f_0$&12.648594&17.4 $\pm$ 3.5&2.496 $\pm$ 0.200&1.03 $\pm$ 0.35&0.06 $\pm$0.32&0.2\nl
$7f_0$ + $f_m$&12.656821&9.6 $\pm$ 3.5&2.575 $\pm$ 0.362& & & \nl
&&&&&&\nl
$8f_0$ - $f_m$&14.447309&11.0 $\pm$ 3.5&0.931 $\pm$ 0.314& &0.10 $\pm$ 0.42&0.2\nl
$8f_0$&14.455536&9.5 $\pm$ 3.5&1.342 $\pm$ 0.365&2.42 $\pm$ 1.00&0.36 $\pm$ 0.34&1.0\nl
$8f_0$ + $f_m$&14.463763&12.1 $\pm$ 3.5&1.034 $\pm$ 0.282& & & \nl
\enddata

\tablecomments{The star pulsates with only one frequency, $f_0$ =
1.806942 $\pm$ 0.000008 {\sl d}$^{-1}$ ({\sl P} = 13.28211 $\pm$
0.00006 {\sl hours}). The amplitude of this frequency is modulated
with the Blazhko frequency of $f_m$ = 0.00823 $\pm$ 0.00004 {\sl d}$^{-1}$
({\sl P}$_{Blazhko}$ = 121.6 $\pm$ 0.6 {\sl d}). Other parameters
are the same as Table \ref{freq-anal-82.8410.55} with
where $n$ = 8 and $t_0$ = HJD 2449656.8470.}

\end{deluxetable}

\newpage
\begin{deluxetable}{lrrrrrc}
\footnotesize
\tablecaption{Frequency Solution for 5.5376.3686
\label{freq-anal-5.5376.3686}}
\tablewidth{0pt}
\tablehead{
&
\colhead{Frequency} &
\colhead{A} &
\colhead{$\varphi$} &
\colhead{${{A_{+1} + A_{-1}} \over {A_0}}$} &
\colhead{$\varphi_{+} - \varphi_{-}$} &
\colhead{ ${{\varphi_{+} - \varphi_{-}} \over {\sigma}}$}\\
&
\colhead{{\sl (d$^{-1}$)}}&
\colhead{{\sl (mmag)}}&
\colhead{{\sl (rad)}}&
&
\colhead{ $\varphi - { {\varphi_+ + \varphi_-}\over{2} } $}&
\colhead{ ${\varphi - {{\varphi_+ + \varphi_-} \over {2}}} \over {\sigma}$ }
}
\startdata
$f_0$ - $f_m$&1.830195&79.9 $\pm$ 5.4&-0.59 $\pm$ 0.07& &-0.01 $\pm$ 0.09&0.0\nl
$f_0$&1.853549&296.6 $\pm$ 5.4&-0.89 $\pm$ 0.02&0.56 $\pm$ 0.03 &-0.30 $\pm$ 0.05&-6.4\nl
$f_0$ + $f_m$&1.876904&87.2 $\pm$ 5.4&-0.60 $\pm$ 0.06& & & \nl
&&&&&&\nl
$2f_0$ - $f_m$&3.683744&24.0 $\pm$ 5.4&2.17 $\pm$ 0.22& &-0.02 $\pm$ 0.24&-0.1\nl
$2f_0$&3.707098&138.5 $\pm$ 5.4&2.09 $\pm$ 0.04&0.58 $\pm$ 0.06&-0.07 $\pm$ 0.12&-0.6\nl
$2f_0$ + $f_m$&3.730453&56.6 $\pm$ 5.4&2.15 $\pm$ 0.10& & & \nl
&&&&&&\nl
$3f_0$ - $f_m$&5.537293&10.6 $\pm$ 5.4&-0.65 $\pm$ 0.51& &-0.24 $\pm$ 0.53&-0.4\nl
$3f_0$&5.560647&88.1 $\pm$ 5.4&-1.00 $\pm$ 0.06&0.52 $\pm$ 0.09&-0.23 $\pm$ 0.27&-0.9\nl
$3f_0$ + $f_m$&5.584002&35.5 $\pm$ 5.4&-0.89 $\pm$ 0.15& & & \nl
&&&&&&\nl
$4f_0$ - $f_m$&7.390842&27.0 $\pm$ 5.4&2.74 $\pm$ 0.20& &-0.26 $\pm$ 0.27&-0.9\nl
$4f_0$&7.414196&49.8 $\pm$ 5.4&2.02 $\pm$ 0.11&1.13 $\pm$ 0.20&-0.59 $\pm$ 0.15&-4.0\nl
$4f_0$ + $f_m$&7.437551&29.1 $\pm$ 5.4&2.48 $\pm$ 0.19& & & \nl
&&&&&&\nl
$5f_0$ - $f_m$&9.244391&6.1 $\pm$ 5.4&-0.38 $\pm$ 0.88& &-0.65 $\pm$ 0.90&-0.7\nl
$5f_0$&9.267745&37.8 $\pm$ 5.4&-0.96 $\pm$ 0.14&0.91 $\pm$ 0.24&-0.25 $\pm$ 0.47&-0.5\nl
$5f_0$ + $f_m$&9.291100&28.4 $\pm$ 5.4&-1.03 $\pm$ 0.19& & & \nl
&&&&&&\nl
$6f_0$ - $f_m$&11.097940&15.2 $\pm$ 5.4&2.59 $\pm$ 0.36& &-0.12 $\pm$ 0.43&-0.3\nl
$6f_0$&11.121294&26.3 $\pm$ 5.4&2.55 $\pm$ 0.20&1.45 $\pm$ 0.42&0.02 $\pm$ 0.25&0.1\nl
$6f_0$ + $f_m$&11.144649&22.9 $\pm$ 5.4&2.47 $\pm$ 0.24& & & \nl
\enddata

\tablecomments{The star pulsates with only one frequency, $f_0$ =
1.853549 $\pm$ 0.000009 {\sl d}$^{-1}$ ({\sl P} = 12.94813 $\pm$
0.00006 {\sl hours}). The amplitude of this frequency is modulated
with the Blazhko frequency of $f_m$ = 0.02335 $\pm$ 0.00002 {\sl d}$^{-1}$
({\sl P}$_{Blazhko}$ = 42.82 $\pm$ 0.04 {\sl d}). Other parameters
are the same as Table \ref{freq-anal-82.8410.55} with
where $n$ = 6 and $t_0$ = HJD 2449891.9649.}

\end{deluxetable}


\clearpage
\begin{figure}
\figurenum{1}
\plotone{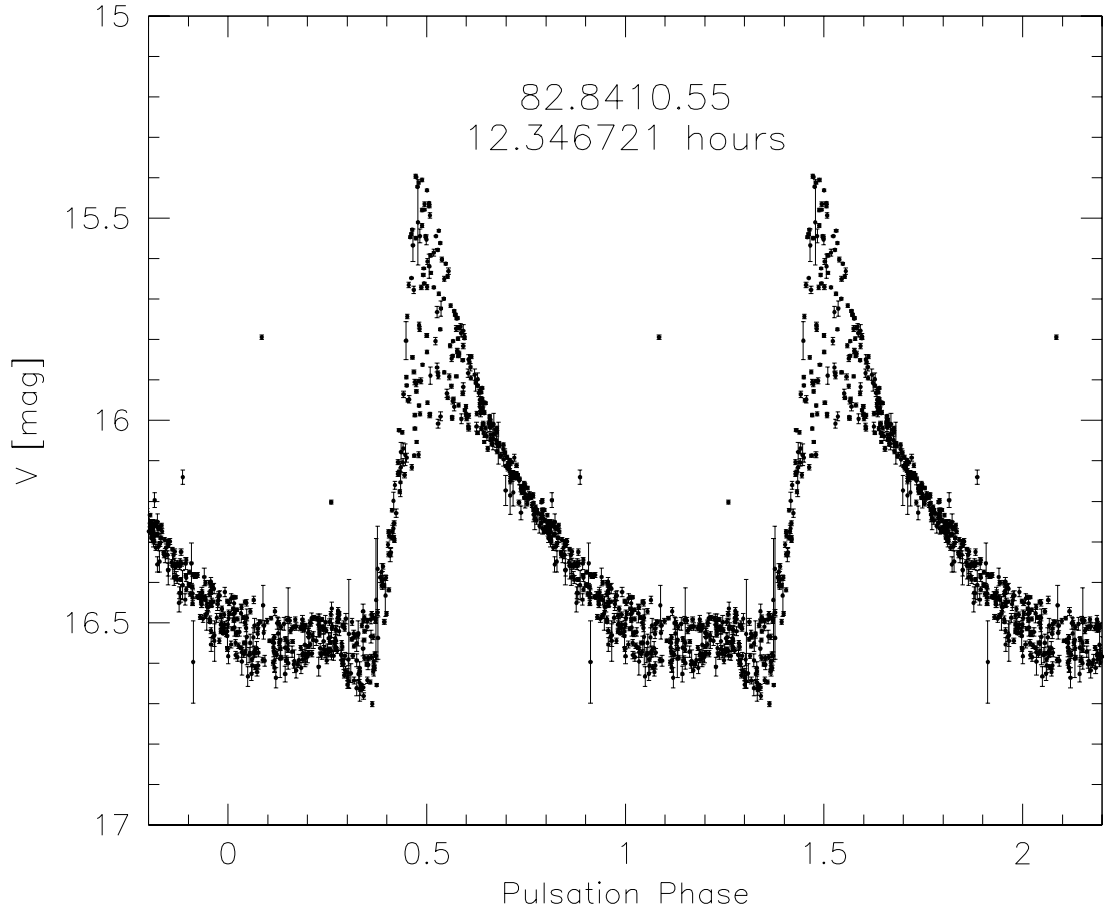}
\caption{
This figure shows all of our data transformed to V for 
the star 82.8410.55 phased with the pulsation period of 12.346721 
$\pm$ 0.00038 {\sl hours}. The amplitude modulation occurs primarily 
at maximum. The tight fit of the rising branch of the lightcurve 
shows the lack of phase modulation over the Blazhko cycle which is 
85.9 $\pm$ 0.2 {\sl d} in this star. All points are plotted twice
in phase to reveal continuity.
}
\end{figure}

\clearpage
\begin{figure}
\figurenum{2}
\plotone{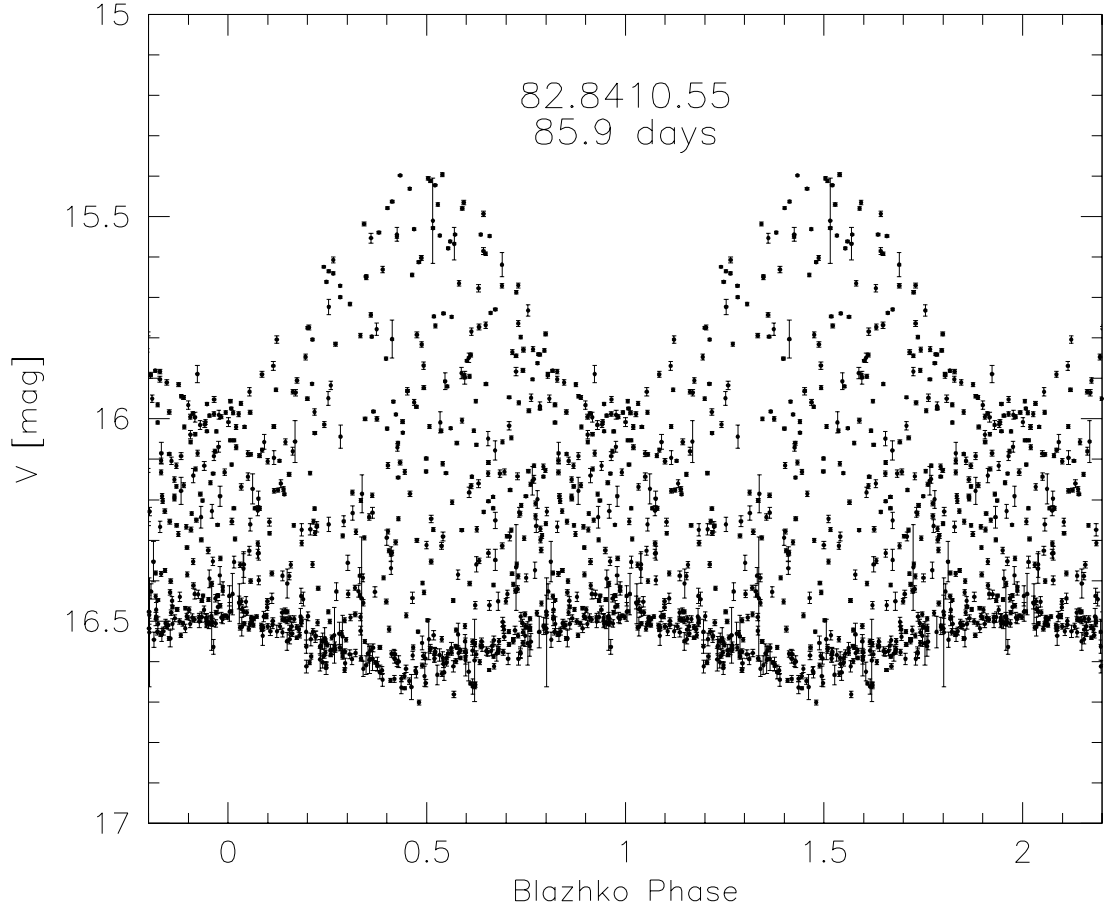}
\caption{
This figure plots the same data as Figure 1,
but now phased with the 85.9 $\pm$ 0.2 {\sl d} Blazhko period. This 
shows clearly for the first time that the amplitude modulation is much 
larger at maximum than at minimum. The physics of this is, as yet, 
unexplained.
}
\end{figure}

\clearpage
\begin{figure}
\figurenum{3}
\plotone{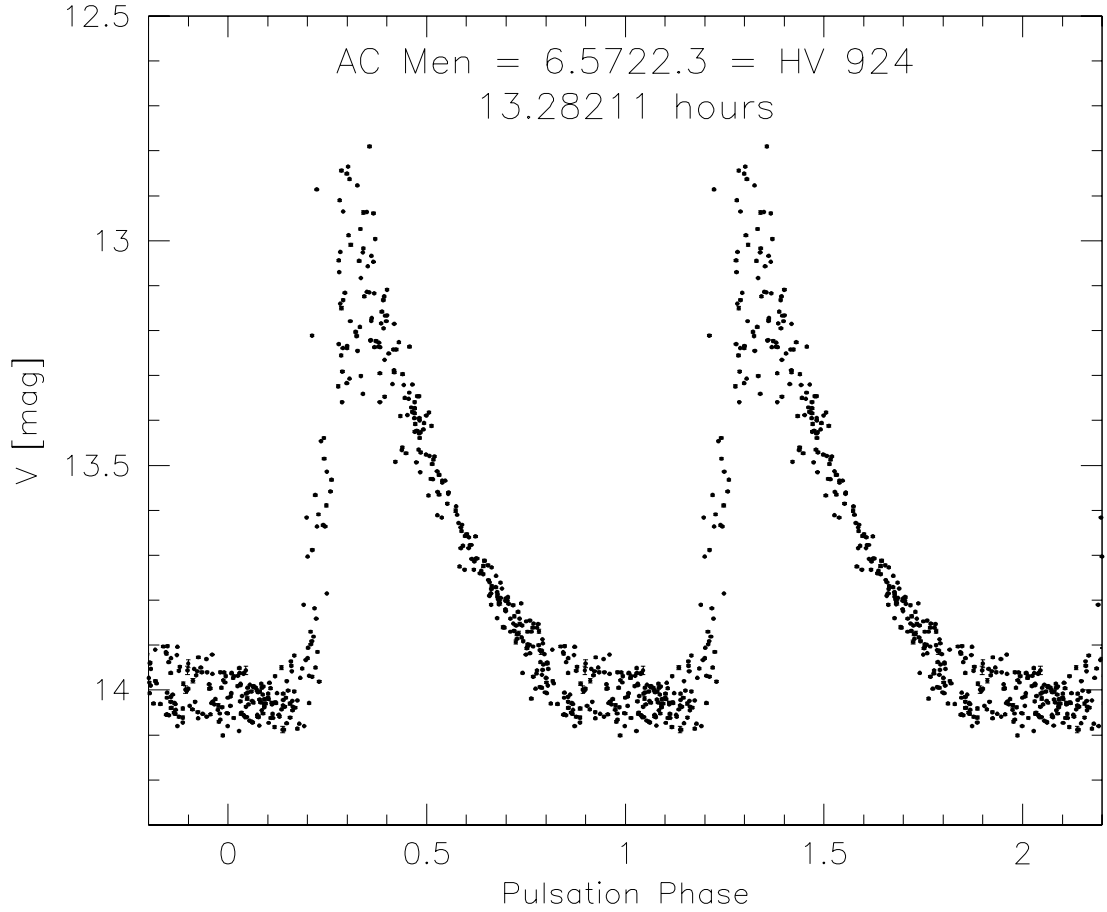}
\caption{
This figure shows all of our data transformed to V for AC Men (6.5722.3)
phased with the pulsation period of 13.28211 $\pm$ 0.00006 {\sl hours}. 
As for star 82.8410.55, the amplitude modulation occurs primarily at 
maximum.  The tight fit of the rising branch of the lightcurve shows 
the lack of phase modulation over the Blazhko cycle which is 121.6 $\pm$ 
0.6 {\sl d} in this star. All points are plotted twice in phase to reveal 
continuity.
}
\end{figure}

\clearpage
\begin{figure}
\figurenum{4}
\plotone{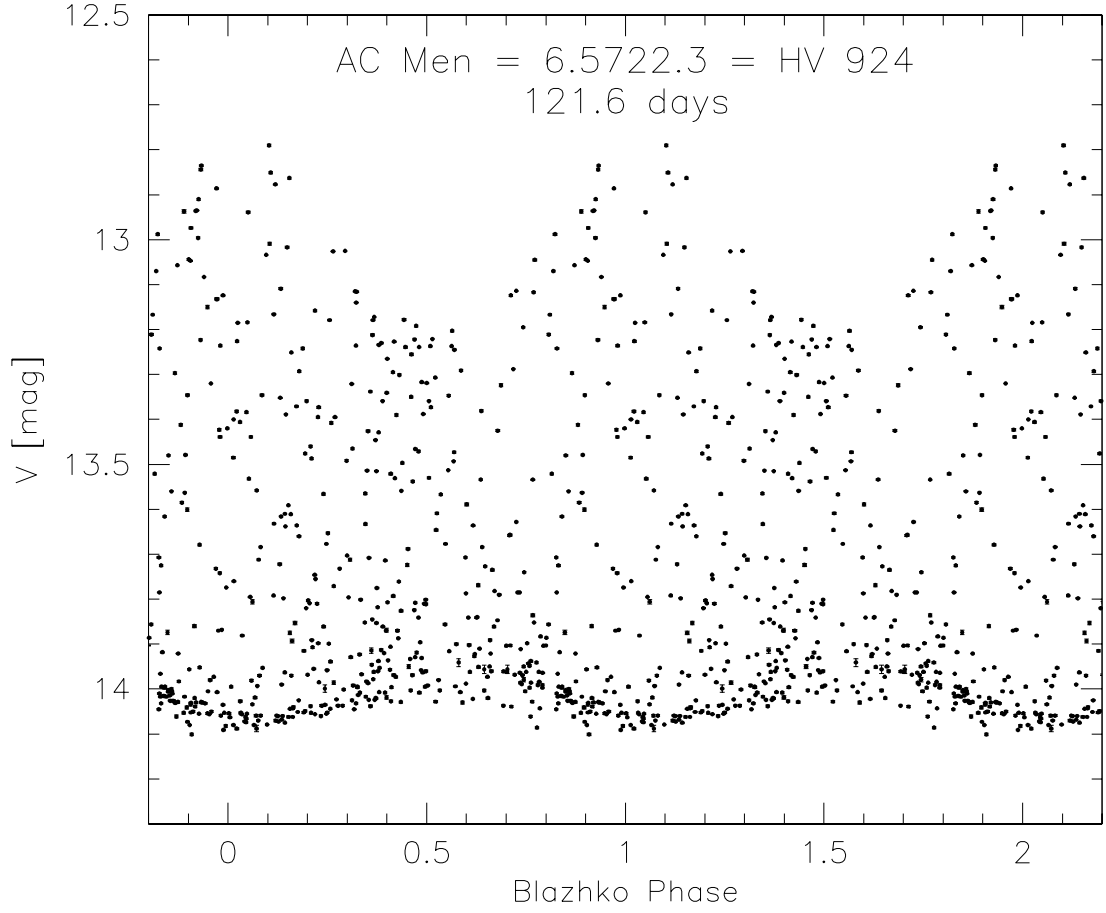}
\caption{\label{phase-6.5722.3-Blazhko}
This figure plots the same data as Figure 3,
but now phased with the 121.6 $\pm$ 0.6 {\sl d} Blazhko period.
}
\end{figure}

\clearpage
\begin{figure}
\figurenum{5}
\plotone{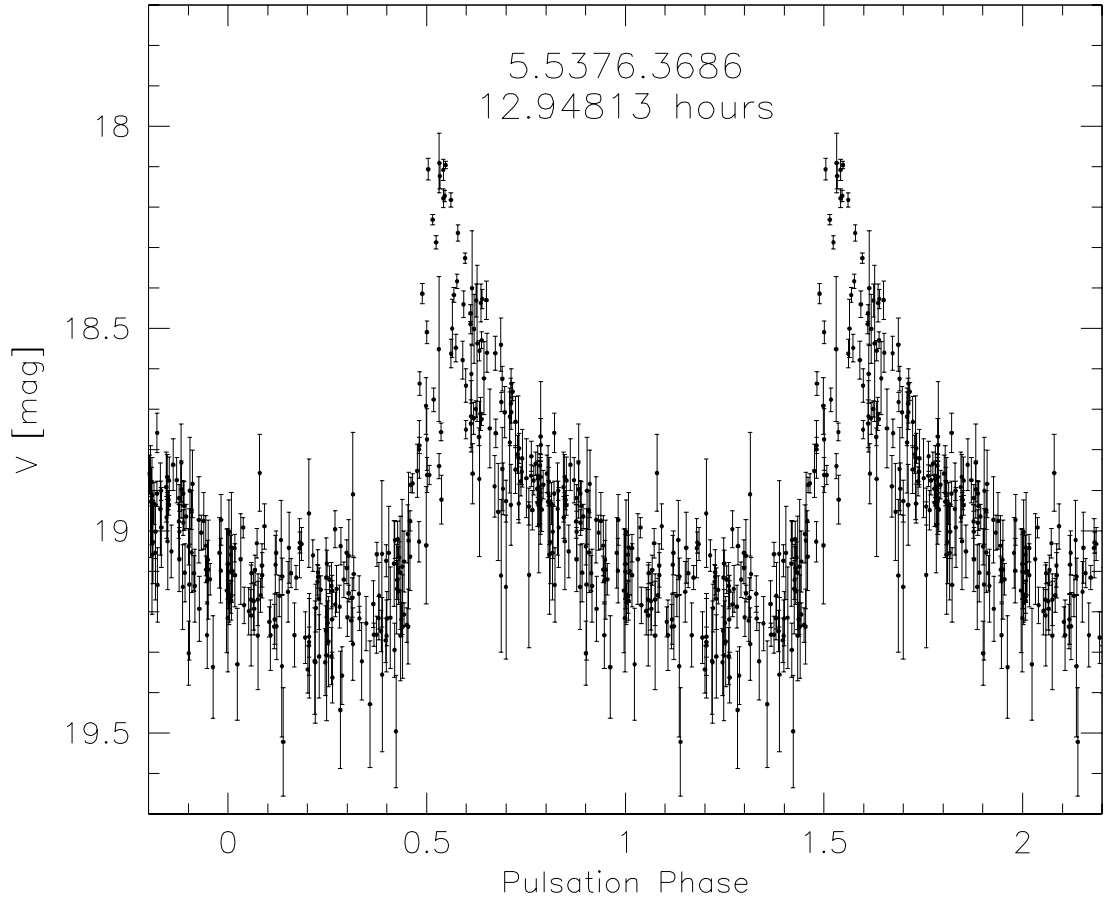}
\caption{
This figure shows all of our data transformed to V for star 5.5736.3686 
phased with the pulsation period of 12.94813 $\pm$ 0.00006 {\sl hours}.
}
\end{figure}

\clearpage
\begin{figure}
\figurenum{6}
\plotone{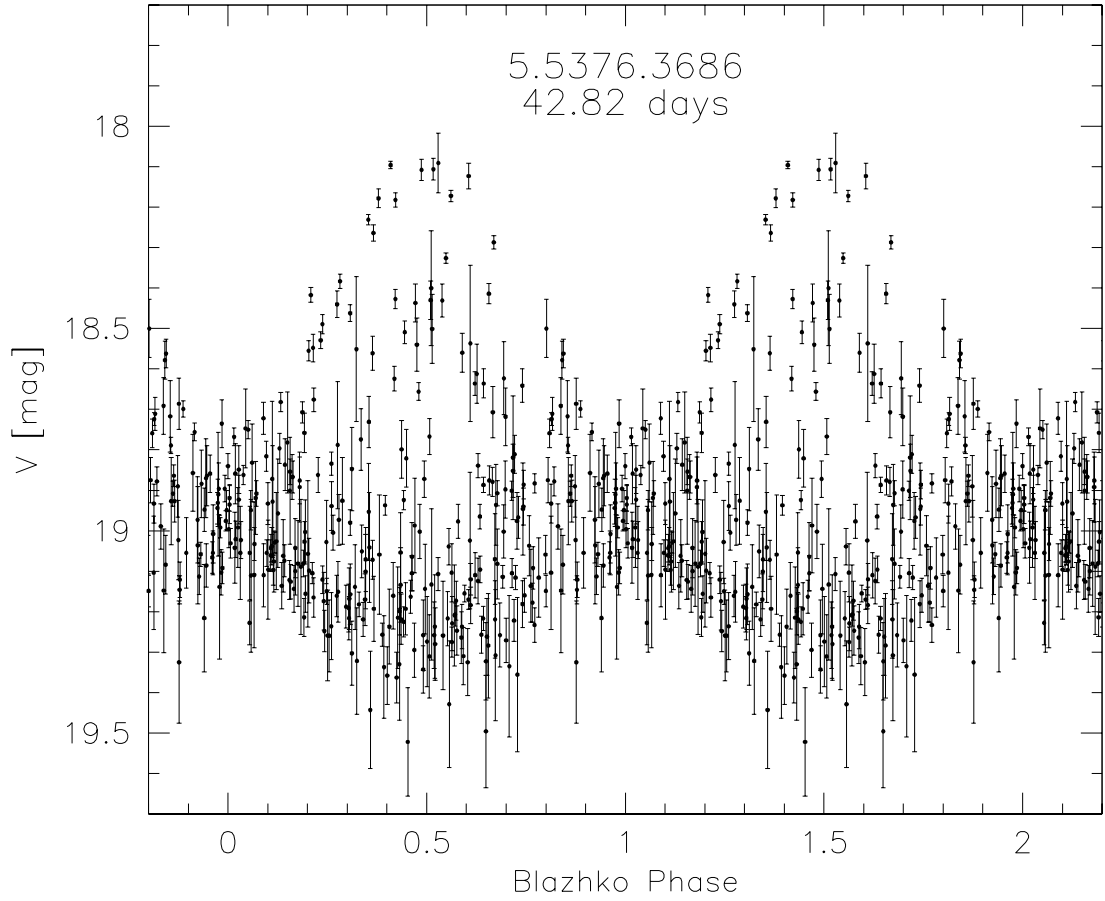}
\caption{
This figure plots the same data as Figure 5,
but now phased with the 42.82 $\pm$ 0.04 {\sl d} Blazhko period.
}
\end{figure}

\clearpage
\begin{figure}
\figurenum{7}
\plotone{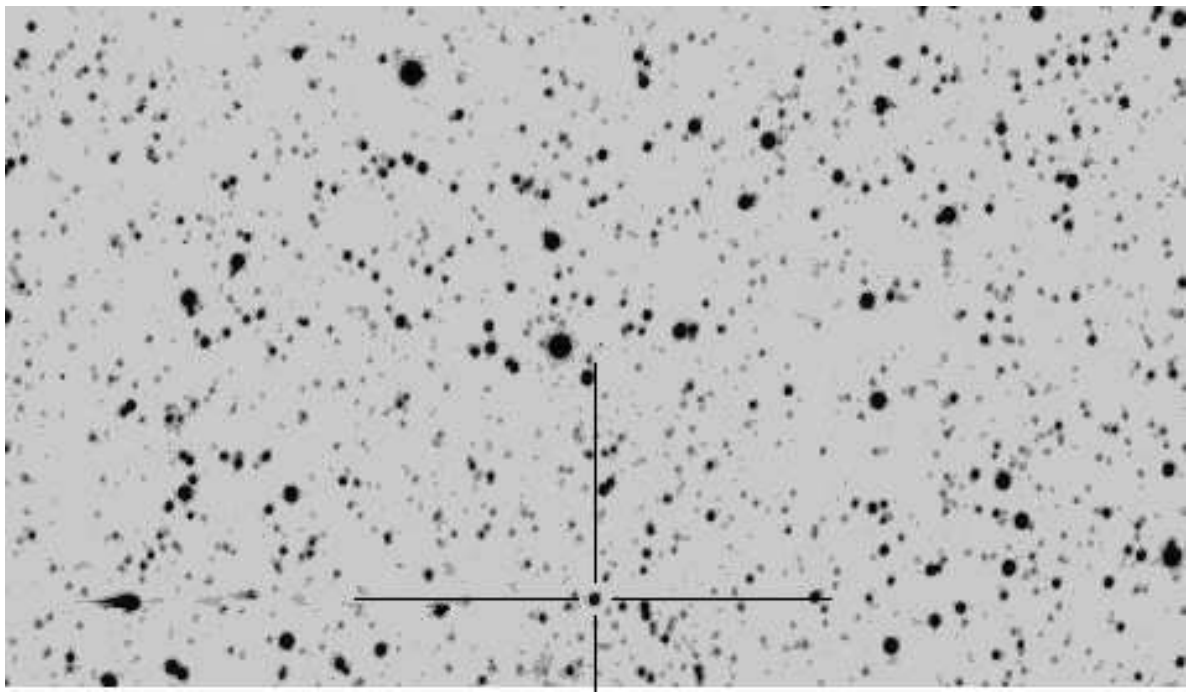}
\caption{This finder chart for the field Blazkho RR Lyrae star
82.8410.55 was obtained in the `red' MACHO bandpass.
The star is indicated by the crosshairs and is located
at equinox J2000.0 coordinates $\alpha$ 05:32:19.0 and
$\delta$ -68:51:52. North is up, east is to the left.
The chart is 3 {\sl arcmin} in its longest dimension.
}
\end{figure}

\clearpage
\begin{figure}
\figurenum{8}
\plotone{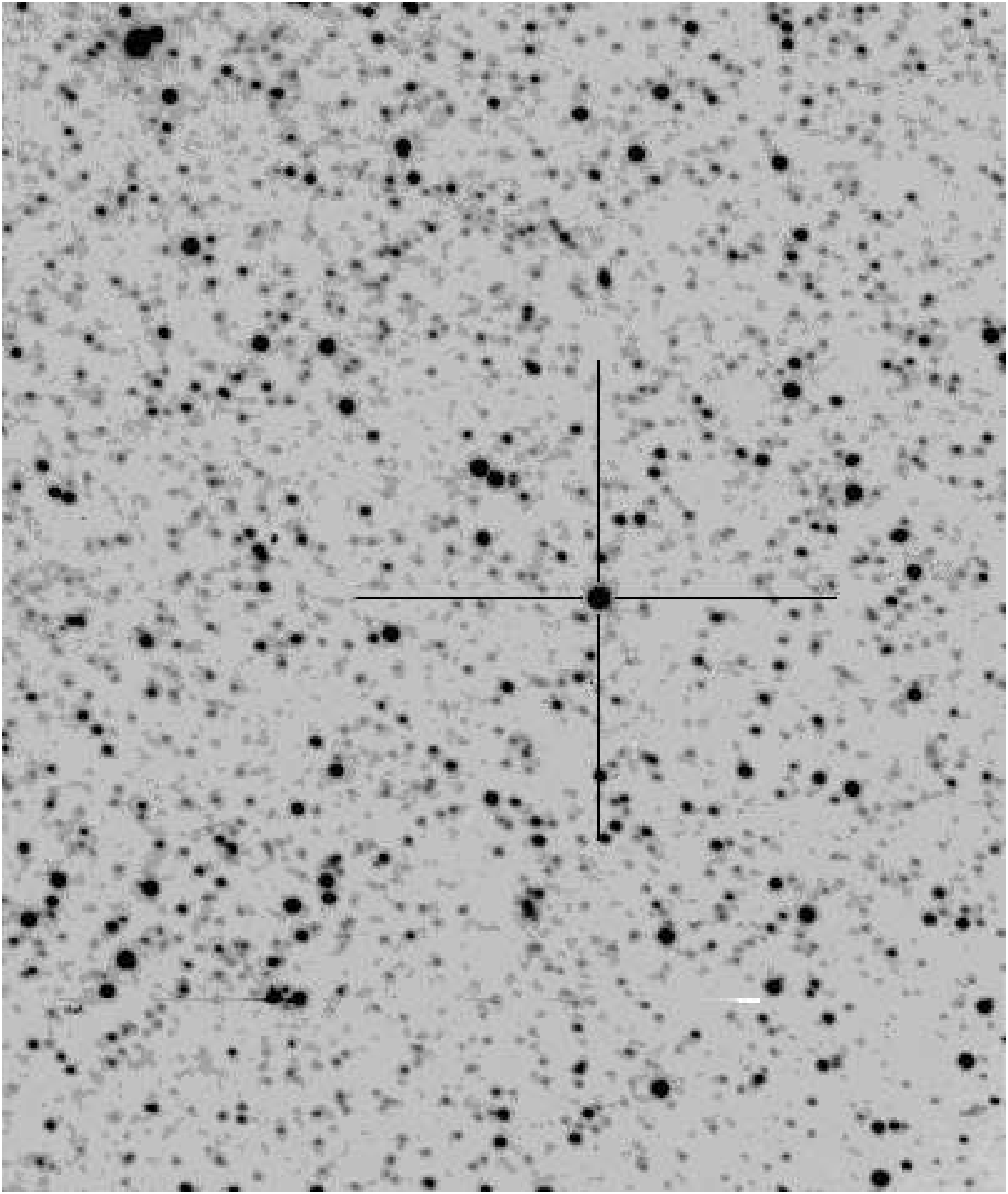}
\caption{This finder chart for the field Blazkho RR Lyrae star
AC Men (6.5722.3) was obtained in the `red' MACHO bandpass.
The star is indicated by the crosshairs and is located
at equinox J2000.0 coordinates $\alpha$ 05:15:50.1 and
$\delta$ -70:36:47. North is up, east is to the left.
The chart is 3 {\sl arcmin} in its longest dimension.
}
\end{figure}

\clearpage
\begin{figure}
\figurenum{9}
\plotone{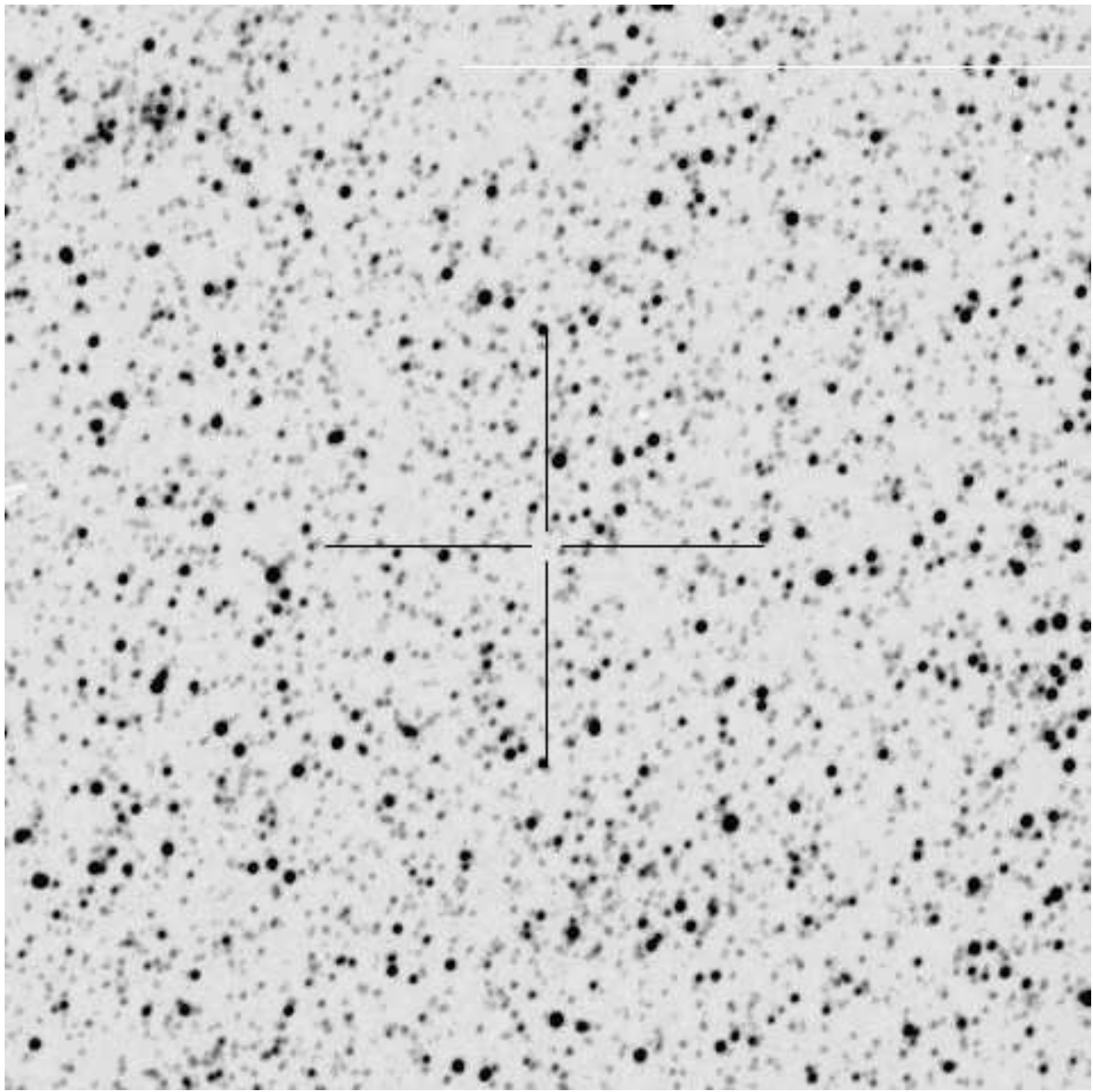}
\caption{This finder chart for the LMC Blazkho RR Lyrae star
5.5376.3686 was obtained in the `red' MACHO bandpass.
The star is indicated by the crosshairs and is located
at equinox J2000.0 coordinates $\alpha$ 05:13:33.6 and
$\delta$ -69:26:21. North is up, east is to the left.
The chart is 3 {\sl arcmin} in its longest dimension.
}
\end{figure}

\end{document}